\documentstyle[12pt,aaspp4]{article}

\received{1997 Sept 22}
\accepted{1997 Nov 18}

\lefthead{Hickson}
\righthead{Measurement of Unresolved Binaries}

\begin{document}

\title{A Technique for Photometric Detection and Measurement of Unresolved
Binary Systems}

\author{Paul Hickson}

\affil{Department of Physics and Astronomy, University of British 
	Columbia, 2219 Main Mall, Vancouver, BC V6T 1Z4, Canada}



\begin{abstract}
A technique is described for the detection and measurement
of close binary systems whose images are unresolved. The method is
based on analysis of the moment of inertia tensor of the image, 
from which the product of the binary flux ratio and square of the angular 
separation may be determined. Intrinsic asymmetries of the  point-spread 
function are removed by comparison with the image of a reference star. 
Multiple exposures may be used to increase the signal-to-noise ratio 
without need of image alignment. An example is given of a simulated
measurement of the dwarf carbon star system G77-61.
\end{abstract}


\section{INTRODUCTION}

There are a number of astrophysical situations where one needs to measure
either the position or the flux of a faint unresolved object which is
located close to a bright star. Obvious examples are the search for
extra-solar planets and low-luminosity companions to bright stars,
and the study of close binary systems. If the flux ratio between the 
primary and secondary object is very large, coronographic techniques may 
be employed, but this is can only be effective if the separation of 
the images of the primary and secondary objects is substantially larger 
than the radius of the point-spread function (PSF). For small separations
alternate techniques are needed.

When the separation between the binary components is comparable to, or
smaller than, the radius of the PSF, the image of the system will be
elongated to some degree. This paper describes a technique which uses
the image elongation to obtain information about the photometric and 
astrometric parameters of the system.

\section{THE ALGORITHM}

There are a several conditions that are desirable in any technique
which aims to measure image elongation. Some important ones are:
\begin{enumerate}
  \item{invariance to the location of the PSF}
  \item{invariance to the orientation of the detector}
  \item{insensitivity to the shape of the PSF}
  \item{optimization of the signal-to-noise ratio}
  \item{ability to co-add images to improve the signal-to-noise ratio}
\end{enumerate}
The method described in this paper satisfies, to a large degree,
all of these conditions.

\subsection{Characterizing Image Elongation}

We begin by reviewing some elementary properties of the moments of the 
light distribution. For a Cartesian coordinate system ${\bf x} = (x^1,x^2)$ 
and intensity distribution $f({\bf x})$, the zeroth, first and second 
moments are defined by
\begin{eqnarray}
  \rho & = & \int f({\bf x}) d^2x \\
  \rho^i & = & \int f({\bf x}) x^i d^2x \\
  \rho^{ij} & = & \int f({\bf x}) x^i x^j d^2x
\end{eqnarray}
where $i = 1,2$ and the function $f({\bf x})$ is presumed to be vanishingly
small at large values of $|{\bf x}|$. The integrals extend over a
sufficiently large range that contributions from the end points are
negligible.

A shift in the position of the image, $f({\bf x}) \rightarrow f({\bf x - a})$,
induces the following changes, which are easily obtained by the
substitution ${\bf y = x - a}$,
\begin{eqnarray}
  \rho & \rightarrow & \rho \\
  \rho^i & \rightarrow & \rho^i + a^i\rho \\
  \rho^{ij} & \rightarrow & \rho^{ij} + a^i\rho^j + a^j\rho^i + a^i a^j \rho
\end{eqnarray}
from which it follows that the inertia tensor
\begin{equation}
I^{ij} = [\rho^{ij} - \rho^i\rho^j/\rho]/\rho
\end{equation}
is invariant under translations of the image.

Under rotations of the coordinate system, $x^{\prime i} = \Lambda^i_{~j} x^j$
(a summation over repeated indices is implied),
it is evident from the definition that $\rho$ and $\rho^i$ transform as
a scalar and vector respectively, and that both $\rho^{ij}$ and $I^{ij}$
transform as second-rank tensors, eg.
\begin{equation}
  I^{\prime ij} = \Lambda^i_{~k}\Lambda^j_{~l} I^{kl}~.
\end{equation}
Where there is no ambiguity, we may omit the indices on vectors, tensors 
and matrices and use bold face type to distinguish them from scalars
quantities.

\subsection{Moments of a Binary Image}

Let $\hat{f}({\bf x})$ be proportional to the PSF. We chose the proportionality
constant and the origin of the ${\bf x}$ coordinate system in such a way that
the zeroth moment is unity and the first moments vanish, ie
\begin{eqnarray}
  \hat{\rho} & = & 1 \\
  \hat{\rho}^i & = & 0
\end{eqnarray}
where the symbol `~$\hat{}$~' indicates that these are normalized moments of 
the PSF. The second moments are generally nonzero and contain information 
about the size and shape of the PSF.

Now consider an image of the form
\begin{equation}
  f({\bf x}) = \hat{f}({\bf x}+{\bf a}/2)+b\hat{f}({\bf x}-{\bf a}/2)
\end{equation}
where $b \leq 1$. This corresponds to a binary system where the secondary
has a flux $b$, relative to the primary, and a vector separation ${\bf a}$. 
Using Eqns. 1-3 and 7, we obtain the inertia tensor of this image
\begin{equation}
  I^{ij} = \hat{I}^{ij} + {b\over (1 + b)^2}a^ia^j~.
\end{equation}
From this we see that the presence of the secondary component adds a term
to the inertia tensor of the PSF which, to first order, is proportional
to the relative flux times the square of the separation. By subtracting
the components of the inertia tensor of the PSF from that of the binary
system, we obtain this term, which we denote by ${\bf M}$,
\begin{equation}
  M^{ij} \equiv I^{ij} - \hat{I}^{ij} = {b\over (1 + b)^2}a^ia^j~.
\end{equation}
From this, if the separation vector ${\bf a}$ is known, we obtain the relative
flux. In fact, only the magnitude $a$ of the separation vector is 
needed - by taking the trace of the matrix ${\bf M}$ we obtain
\begin{equation}
  Tr({\bf M}) = {ba^2\over (1 + b)^2}~,
\end{equation}
from which $b$ may be determined.

Alternatively, if the relative flux is known, the diagonal terms of ${\bf M}$
give the components of separation vector. Even if $b$ is not known, the 
position angle $\phi$ of the separation vector, with respect to the 
$x^1$ axis, can be found from the ratio of diagonal terms,
\begin{equation}
  \tan\phi = a^2/a^1 = (M^{22}/M^{11})^{1/2}~.
\end{equation}

\subsection{Sampling Effects}

In practice we work with a digital image which consists of a discrete set 
of samples $f_\alpha$
corresponding to pixels located at positions ${\bf x}_\alpha$, normally
on a square grid at intervals $p$. The 
values $f_\alpha$ are proportional to the product
of the intensity and the pixel response functions $q_\alpha({\bf x})$
integrated over the pixel areas. If the response function is the same
for all pixels we have
\begin{eqnarray}
  f_\alpha & = & \int f({\bf x}) q_\alpha({\bf x}) d^2x \\
    & = & \int f({\bf x}) q({\bf x}-{\bf x}_\alpha) d^2x~.
\end{eqnarray}
Thus the values $f_\alpha$ actually sample the image formed by smoothing the
incident intensity with the pixel response function. This effect can 
be incorporated by employing the actual PSF of the sampled image, which 
includes the result of the smoothing.

From the sampling theorem (see, for example, \cite{b78}) the 
function $f({\bf x})$ can be uniquely reconstructed from the samples
$f_\alpha$ only if it contains no spatial frequencies higher than the
Nyquist frequency $1/p$. For a diffraction limited image, at wavelength
$\lambda$, from a telescope of aperture diameter $D$, the pixel spacing 
must be no greater than $\lambda/2D$. The image is then given by
\begin{equation}
  f({\bf x}) = \sum_\alpha f_\alpha {\rm sinc}(x^1 - x^1_\alpha)
    {\rm sinc}(x^2 - x^2_\alpha)
\end{equation}
where ${\rm sinc}(x) = \sin({\pi x})/\pi x$. Substitution of this
equation into Eqns. 1-3 gives the corresponding discrete expressions
\begin{eqnarray}
  \rho & = & \sum_\alpha f_\alpha ~,\\
  \rho^i & = & \sum_\alpha f_\alpha x_\alpha^i~, \\
  \rho^{ij} & = & \sum_\alpha f_\alpha x_\alpha^i x_\alpha^j
\end{eqnarray}
providing that the integrals converge. Eqns. 7, 13-15 and 19-21
then provide a direct path from the observational data to the parameters
$b$ and ${\bf a}$ of the binary system.

While this simple technique seems plausible, in practice it does not
work due to problems of convergence and noise, discussed next. However a 
modification of the technique, developed in Section 2.5 avoids these
difficulties while at the same time optimizing the signal-to-noise ratio.

\subsection{Convergence and Noise}

The above algorithm fails because for real astronomical images the 
integrals in Eqns. 1-3 do not converge. This is a result of the 
PSF of an optical system being a band-limited function. The amplitude
of the light in the PSF is the two-dimensional spatial Fourier transform of 
the optical transfer function (OTF). There is an upper limit to the spatial
frequency at which the OTF can be non-zero, imposed by the finite
size of the telescope entrance pupil. The OTF is the product of the 
instrumental and atmospheric transfer functions with the pupil function,
defined as unity for points within the pupil and zero for points outside.
From the convolution theorem, the amplitude of the PSF is therefore the
convolution of the instrumental and atmospheric response with the Fourier 
transform of the pupil
function. Since the square of the latter function is just the 
diffraction-limited PSF, it follows that the intensity of light in the PSF
cannot fall off faster than the diffraction limit. For a circular
aperture this has the form (eg. \cite{bw80})
\begin{equation}
   \hat{f}({\bf x}) = [2J_1(kRr)/kRr]^2
\end{equation}
where $r = |{\bf x}|$ is measured in radians, $k = 2\pi/\lambda$ is
the wave number, $\lambda$ the wavelength, and $R = D/2$ is the radius of the 
aperture. For large values of its argument, the Bessel function has the 
asymptotic expansion
\begin{equation}
  J_1(x) \rightarrow {1\over\sqrt{\pi x}}(\sin{x} - \cos{x})
\end{equation}
so $\hat{f}$ falls off no faster than $r^{-3}$. Since $d^2x \propto r$,
the integral in Eqn. 3 is ill-defined.

Even more serious is the fact that random noise, present in the image,
causes a rapid divergence of the moments. To see this, recall that
both the photon noise and the read noise is uncorrelated from pixel to
pixel. The presence of such noise adds a fluctuating component to 
$f({\bf x})$ which in turn produces fluctuations in the moments. The 
variance of these fluctuations is the sum of the variances of the individual 
pixel fluctuations, weighted by the {\it squares} of the coordinate terms 
present in Eqns. 2 and 3. Since the read noise is independent of position, 
the variance of all moments increases without limit over the range of 
integration.

\subsection{Weighted Moments}

The solution that we propose for these problems is to apply weight
factors to the intensity values of the image before computing the moments.
In order to preserve the transformation properties of the inertia
tensor, the weights cannot depend on the coordinates, but can only be
a function of $f_\alpha$.
We make the substitution $f_\alpha \rightarrow w_\alpha f_\alpha$,
and ask what choice of $w_\alpha$ minimizes the random error in the
moments. From Eqns. 19-21 we then have
\begin{eqnarray}
  Var(\rho) & = & \sum_\alpha Var(f_\alpha) \\
  Var(\rho^i) & = & \sum_\alpha (x_\alpha^i)^2 Var(f_\alpha) \\
  Var(\rho^{ij}) & = & \sum_\alpha (x_\alpha^i x_\alpha^j)^2 Var(f_\alpha)
\end{eqnarray}
where $Var(x)$ denotes the variance of the random variable $x$.
We seek to minimize the ratios $Var(\rho)/(\rho)^2$, $Var(\rho^i)/(\rho^i)^2$
and $Var(\rho^{ij})/(\rho^{ij})^2$, and so equate to
zero the derivatives of these ratios with respect to $w_\alpha$. This
leads to the result
\begin{equation}
  w_\alpha \propto {f_\alpha\over Var(f_\alpha)}~.
\end{equation}
Two possible dependencies for the noise are $Var(f_\alpha) \propto f_\alpha$
and $Var(f_\alpha) = constant$. The first case corresponds to photon 
(Poisson) noise. It leads to $w_\alpha = constant$ which, as we have seen,
is unacceptable. The second case corresponds to a constant read noise.
It leads to the choice $w_\alpha \propto f_\alpha$. As the inertia
tensor is independent of the normalization of $f_\alpha$, we may choose
the proportionality constant to be unity. This choice of weights
solves the convergence problems as the asymptotic dependence of 
$w_\alpha f_\alpha = f_\alpha^2$ is now proportional to $r^{-6}$.

While the use of weights solves the convergence and noise
problems, it introduces complications in the interpretation of the
results. Our choice of weights is equivalent to performing
the previous (unweighted) analysis on a new image formed by squaring
the intensity values of the original image. Thus Eqn. 11 is replaced by
\begin{equation}
  f({\bf x}) = [\hat{f}({\bf x}+{\bf a}/2)+b\hat{f}({\bf x}-{\bf a}/2)]^2~.
\end{equation}
After some algebra, we obtain the inertia tensor
\begin{equation}
  I^{ij} = \hat{I}^{ij} 
    + {b\over 2}\left\{{(1+b^2)\gamma({\bf a})+2b\over 
    [1+b^2+2b\gamma({\bf a})]^2}\right\}a^ia^j
    + {2b\gamma({\bf a})\over 1+2b\gamma({\bf a})+b^2}
    [\hat{I}^{ij}({\bf a})-\hat{I}^{ij}]
\end{equation}
where
\begin{eqnarray}
  \gamma({\bf a}) & = & {\int \hat{f}({\bf x}+{\bf a}/2)\hat{f}({\bf x}-{\bf a}/2)
    d^2x \over \int \hat{f}^2({\bf x}) d^2x} \\
  \hat{I}^{ij}({\bf a}) & = & {\int \hat{f}({\bf x}+{\bf a}/2)
    \hat{f}({\bf x}-{\bf a}/2)x^ix^j d^2x \over \int \hat{f}^2({\bf x}) d^2x}~.
\end{eqnarray}
This result is similar in form to Eqn. 12 but has an extra term involving
the factor $\hat{I}^{ij}({\bf a}) - \hat{I}^{ij}$. This factor depends on the
shape of the PSF and the separation ${\bf a}$. For any separation, it
can be computed numerically, using the image of the reference star to
estimate the PSF.

The function $\gamma({\bf a})$ approaches unity for small values of $a$ and
zero for large $a$. For a Gaussian PSF it has the simple analytic form
\begin{equation}
  \gamma({\bf a}) = \exp(-a^2/4\sigma^2)
\end{equation}
where $\sigma$ is related to the FWHM $w$ of the PSF by 
$w^2 = 8\ln{2} \sigma^2$. For close binary systems $a/\sigma$ is small
and $\gamma({\bf a})$ is quite close to unity.

Eqns. 29-31, which represent the generalization of Eqn. 12 in the weighted
case, relate the binary system 
parameters to the difference in the inertia tensor components
between the binary and comparison images. 

\subsection{Co-adding Images}

The signal-to-noise ratio that can be achieved in a single image is
limited by the dynamic range of the detector. However, the signal-to-noise
ratio can be increased by the use of multiple exposures. A
high-signal-to-noise-ratio image can be obtained by co-aligning the 
individual images, interpolating and summing the individual intensities.
However, the complexity of this process can be avoided if one
wants only to measure the image ellipticity.

The moments, Eqns. 1-3, are linear in $f$. Because of this, the sum
of the moments of a number of individual images are equivalent to the 
moments of the summed image. Moreover, because ${\bf I}$ is independent
of translations, it is not necessary or desirable to co-align the 
individual images. One simply computes the inertia tensor ${\bf I_\alpha}$
for each image and averages the corresponding components:
\begin{equation}
  I^{ij} = {1\over N}\sum_{\alpha = 1}^N I^{ij}_\alpha~.
\end{equation}
This method is preferable to computing the inertia tensor
for a coadded image because the latter technique is sensitive to 
errors in alignment of the individual images.

\section{PERFORMANCE}

\subsection{Noise Analysis}

The analysis of noise propagation in Equations (7), (19-21), and (28) is 
straight forward.
Considerable simplification results if we make the reasonable assumption
that the PSF is circularly symmetric, at least to first order. For the
limiting cases in which the noise variance per pixel is either constant
(read noise dominated) or proportional to the intensity (photon noise 
dominated) we find
\begin{equation}
  Var(I^{ij}) = {1\over s^2}[(I^{ij})^2 + I_n^{iijj}]
\end{equation}
where $s$ is the total signal-to-noise ratio of the image,
\begin{equation}
  s = \rho/[Var(\rho)]^{1/2}
\end{equation}
and 
\begin{equation}
  I_n^{ijkl} = {\sum (f_\alpha)^{n/2} x_\alpha^i x_\alpha^j x_\alpha^k 
    x_\alpha^l \over \sum (f_\alpha)^{n/2} }
\end{equation}
where $n = 2$ for the read-noise-dominated case and $n = 3$ for the
photon-noise-dominated case.

For the case of diffraction-limited imaging by a circular aperture, the
tensors may be evaluated analytically. Using Eqn. 22 and replacing the
summation by integration in Eqn. 36, and recalling that we are
working with the square of the intensity (Section 2.5), we obtain
\begin{eqnarray}
  I^{11} = I^{22} & = & { 16\int_0^{2\pi}\int_0^\infty [J_1(kRr)/kRr]^4 
  	cos^2(\phi) r^3dr d\phi \over 32\pi\int_0^\infty [J_1(kRr)/kRr]^4 rdr }
  	  \nonumber \\
  	 & = & {1\over 2k^2R^2} \int_0^\infty J_1(x)^4 x^{-1}dx / 
  	   \int_0^\infty J_1(x)^4 x^{-3}dx\nonumber \\
  	 & = & 0.088179 (kR)^{-2}~.
\end{eqnarray}
Similarly,
\begin{eqnarray}
  I_2^{1111} = I_2^{2222} = I_2^{1122}
  	 & = & {3\over 8k^4R^4} \int_0^\infty J_1(x)^4 xdx / 
  	   \int_0^\infty J_1(x)^4 x^{-3}dx\nonumber \\
  	 & = & 9.76197 (kR)^{-4}~,\\
  I_3^{1111} = I_3^{2222} = I_3^{1122}
  	 & = & {3\over 8k^4R^4} \int_0^\infty J_1(x)^6 x^{-1}dx / 
  	   \int_0^\infty J_1(x)^6 x^{-5}dx\nonumber \\
  	 & = & 1.01967 (kr)^{-4}
\end{eqnarray}
which gives
\begin{eqnarray}
  Var(I^{11}) = Var(I^{22}) & = & \left\{\begin{array}{c}
    10.53953 \\ 1.797219 \end{array}\right\} {1\over s^2}
    \left[{\lambda\over 2\pi R}\right]^4 \\
  Var(I^{12}) & = & \left\{\begin{array}{c} 9.76197 \\ 1.01967 \end{array}
    \right\} {1\over s^2} \left[{\lambda\over 2\pi R}\right]^4
\end{eqnarray}
where the upper numerical values refer to the read-noise-dominated
case and the lower values refer to the photon-noise-dominated case.

From Eqn. 29 it is evident that there is no simple relation between
the binary system parameters and the inertia tensor. Simulations
indicate that the last term in this equation typically contributes
about 1/3 of the ellipticity signal. If we assume that the variance 
in the image of the comparison star image is comparable to that of 
the binary system, we expect that
\begin{equation}
  Var(ba^ia^j) \sim 10~ Var(I^{ij}) ~.
\end{equation}
Eqns. 40-42 can be used to estimate the signal-to-noise ratio that would 
be required for any desired measurement of the binary system parameters. 
For example, if we assume that the magnitude $a$ of the separation 
vector is known, the error in the measured flux ratio becomes, for the 
photon-noise-limited case,
\begin{equation}
  \sigma(b) \equiv [Var(b)]^{1/2} \sim {1\over s} \left[{\lambda \over
  aD}\right]^2~.
\end{equation}
Thus, in order to detect and measure, to 10\% accuracy, a binary which has 
flux ratio 0.01 and separation comparable to the size of the
diffraction-limited PSF, $a = \lambda/D$, we need $\sigma(b) = 0.001$
which implies a total signal-to-noise ratio of order $s \sim 10^3$.

\subsection{Simulations}

In order to verify this analysis and evaluate the performance of the method,
numerical simulations were performed. In a series of Monte-Carlo runs, 
artificial images were created of both a binary system and a reference star,
by placing a suitably-scaled PSF image at the location of each component 
and then adding random noise. For each run the program estimated the 
desired parameters. The mean values and standard errors of the parameters 
were then determined for the set of runs.

As an example, consider the dwarf carbon star system G77-61 (\cite{d.77}, 
\cite{d.86}). The secondary is believed to be a population-II white dwarf, 
and a measure of its luminosity can provide a lower limit to the age of
the universe (\cite{r.97}). In the near infrared, the white dwarf is 
expected to be 4 to 5 magnitudes fainter than the primary star, and 
the maximum separation of the system is estimated to be 0.035 arcsec.
For a hypothetical observation using the Hubble Space Telescope (HST) and 
NIC1 infrared camera, The pixel size is 0.043 arcsec and the
aperture diameter is 2.4 m. Thus, to avoid aliasing, we must 
use filters which block wavelengths below 1.0 um. The flux ratio
is expected to be 0.009 and 0.021 at a wavelength of 1.10 um and 1.55 um 
respectively. We regard the separation as known and wish to measure the 
relative flux from the observations. 

The results of simulations of this system are presented in Figs. 1 and 2 
and Table 1.
These show the estimated value of the flux ratio $b$, and its statistical
error, as a function of the total signal-to-noise ratio $s$, for two different
wavelengths. At low values of $s$, noise fluctuations dominate, producing 
a substantial random elongation of the image. As a result, both $b$ and
its error are large.  As $s$ increases, the noise fluctuations
decrease and the image becomes more circular -- both $b$ and its error
decrease. At $s \simeq 10^3$, the binary nature of the star
prevents further circularization of the image, and $b$
stabilizes at the correct flux ratio. When $s \simeq 10^4$, the relative 
error of $b$ has dropped to about 0.1. From this we can conclude that
in order to measure the flux ratio with an error of 10\%, a total
signal-to-noise ratio of order $10^4$ is required. This is much greater 
than can be obtained in a single image, but could be achieved by coadding
many exposures. The exposure time for the individual images should be 
chosen to maximize the signal-to-noise ratio, while remaining in the linear 
region of the detector. As this results in a constraint on the maximum 
flux in any individual pixel, it may be more convenient to work with the 
peak signal-to-noise ratio $s_p$, ie. that of the central pixel, rather 
than the total signal-to-noise ratio $s$. For the photon-noise-limited 
case, the two quantities are related by the square root of the fraction 
of the total light contained within the central pixel
\begin{equation}
  s_p = s~{p R \sqrt{\pi}\over \lambda}~.
\end{equation}
Both total and peak signal-to-noise ratios are listed in Table 1.

The results of the simulation show that the analytic error estimate
of Eqn. 43 is reasonably accurate. For the 1.1-um simulation, 
this equation predicts that a total signal-to-noise ratio $s \sim 8000$
would be required for a 10\% measurement of $b$ which, given the
approximation in estimating the Variance from Eqn. 28, is in
quite good agreement with the value $s \sim 10000$ indicated by the 
simulation.

\section{DISCUSSION}

We have described a new technique for the detection and measurement
of unresolved binary systems based on the elongation of their images, 
in comparison to images of refererence stars. The method can only work 
if variations the shape of the PSF, between the binary and reference 
observations, are smaller than the elongation to be measured. The 
analysis and simulations show what signal-to-noise ratio must be
obtained in order to reduce the random errors to any desired level.
However, one must also consider systematic errors that may not be
removed by the differential measurement technique. These include
optical aberrations, guiding errors, and any other effects of this
nature. Aberrations can be minimized by positioning the reference
star and binary at precisely the same location on the detector.
Guiding errors which are the same for both reference and target
observations will be removed by the analysis, but any variations
will contaminate the signal. Such effects should be characterized
and understood before observations are attempted. For HST,
guiding errors are reported to be of order 0.001 arcsec on an
individual exposure. If these are random, they can be reduced to
sufficiently small levels by obtaining many exposures.

Ideally, it would be best to observe the target and reference stars 
simultaneously, if their separation is sufficiently small that both
images can simultaneously fit on the detector. In this case asymmetries 
due to guiding errors should be the same for both images and will be 
removed by the analysis. in order to minimize optical aberrations, 
the stars should be placed on opposite sides of the optical field 
center, and equidistant from it. To further reduce systematic effects, 
the telescope should be rotated axially by $180\deg$, if possible, to 
interchange the positions of target and reference stars. By exposing
for equal times in the two configurations, the systematic 
component of the PSF will then be identical for both target and 
reference images.

It should be possible to apply this technique to images produced by
adaptive optics systems on ground-based telescopes. The high-resolution 
provided by such systems facilitates the detection of close
binaries, but great care will be needed to minimize variations in
the PSF between the binary system and the comparison star. Because 
the performance of adaptive optics systems depends the brightness of 
the reference star and on the atmospheric seeing, the binary system
and calibration star need to be well-matched in both magnitude and
sky position. We hope to test the feasibility of the method
in the near future by means of adaptive-optics observations with the
Canada-France-Hawaii Telescope.

\acknowledgments

I am grateful to Harvey Richer for bringing the problem of the measurement
of G77-61 to my attention, and for several interesting discussions.
An anonymous referee provided helpful comments and emphasized the 
importance of matching the comparison and target stars.
This work was supported by grants from Natural Sciences and Engineering 
Research Council of Canada. 

\clearpage

\begin{deluxetable}{lrrrrrr}
\tablenum{1}
\tablewidth{0pt}
\tablecaption{Simulations of the G77-61 Binary System}
\tablehead{ & 
\colhead{-----------} & 
\colhead{1.10 um} & 
\colhead{-----------} & 
\colhead{-----------} & 
\colhead{1.55 um} &
\colhead{-----------} 
\nl
\colhead{$s$} & 
\colhead{$s_p$} &
\colhead{$<b>$} & 
\colhead{$\sigma(b)$} & 
\colhead{$s_p$} &
\colhead{$<b>$} & 
\colhead{$\sigma(b)$}
}
\startdata
   10 &     4.03 & 0.507294 & 0.486904 &     2.86 & 0.472412 & 0.494915 \nl
   22 &     8.68 & 0.428661 & 0.435401 &     6.16 & 0.460700 & 0.476939 \nl
   46 &    18.71 & 0.136120 & 0.198945 &    13.28 & 0.307730 & 0.391819 \nl
  100 &    40.31 & 0.049294 & 0.065857 &    28.61 & 0.125452 & 0.184577 \nl
  215 &    86.84 & 0.023317 & 0.027893 &    61.63 & 0.054850 & 0.067542 \nl
  464 &   187.10 & 0.013118 & 0.013616 &   132.78 & 0.030990 & 0.031048 \nl
 1000 &   403.09 & 0.009182 & 0.007273 &   286.07 & 0.022006 & 0.016621 \nl
 2154 &   868.43 & 0.008086 & 0.003830 &   616.30 & 0.019957 & 0.008461 \nl
 4642 &  1871.00 & 0.007879 & 0.001808 &  1327.81 & 0.019566 & 0.003946 \nl
10000 &  4030.94 & 0.007801 & 0.000839 &  2860.67 & 0.019414 & 0.001832 \nl
21544 &  8684.26 & 0.007766 & 0.000389 &  6163.02 & 0.019347 & 0.000850 \nl
46416 & 18710.01 & 0.007750 & 0.000181 & 13278.07 & 0.019317 & 0.000395 \nl
100000 & 40309.41 & 0.007743 & 0.000084 & 28606.68 & 0.019304 & 0.000183 \nl
\hline\vspace{-18pt}
\enddata
\end{deluxetable}

\clearpage

\noindent Figure Captions

\figcaption[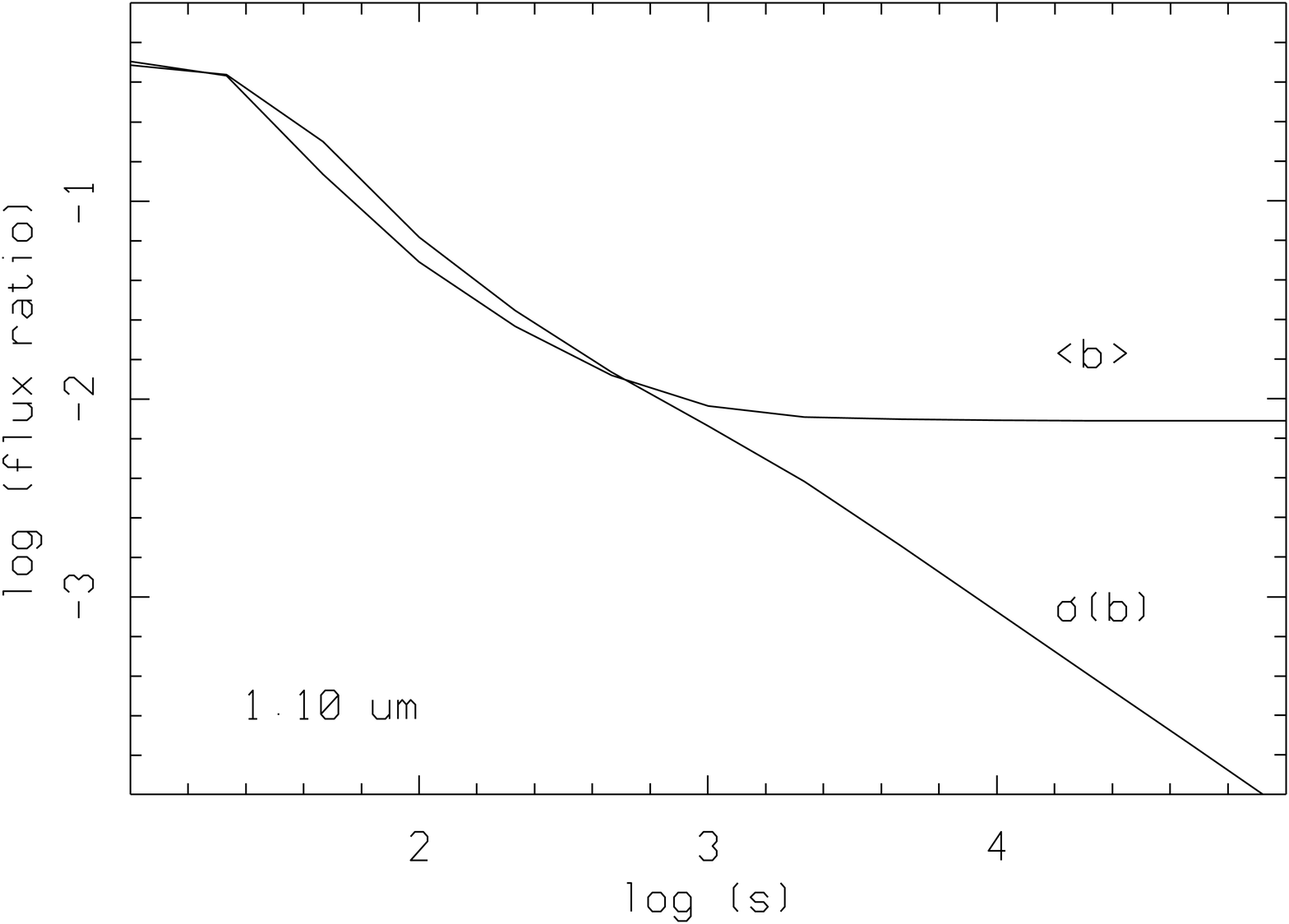]{Detecting a simulated binary system. The curves show the
estimated flux ratio of the binary, and its standard error, as a
function of the total signal-to-noise ratio in the image. The binary
system is assumed to have a true flux ratio of 0.00909 and a separation 
of 0.035 arcsec. The simulation is for a 2.4-meter telescope at a 
wavelength of 1.10 um.\label{fig1}}

\figcaption[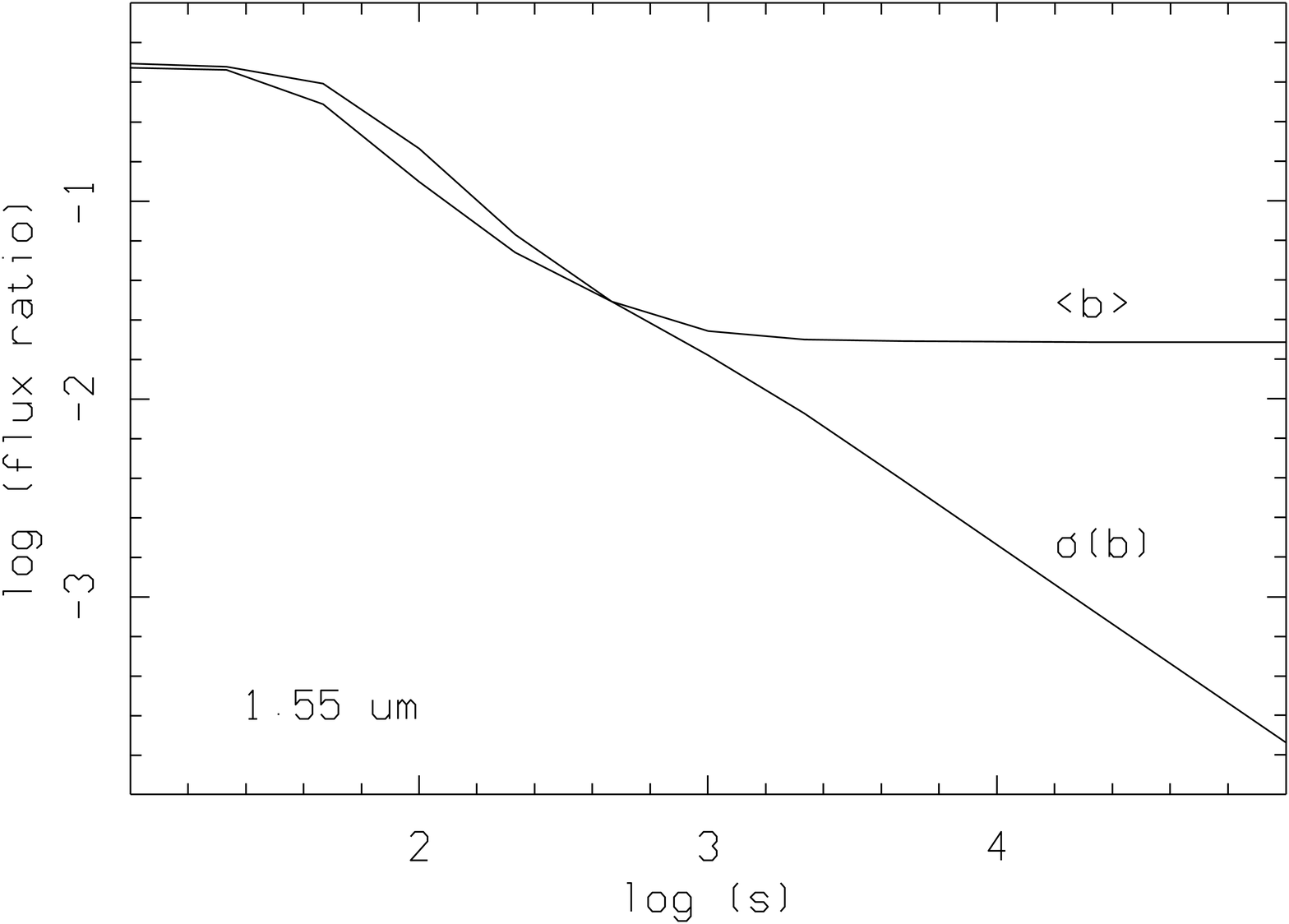]{This plot is similar to Fig. 1, but for a flux ratio of 0.02083
and a wavelength of 1.55 um.\label{fig2}}

\clearpage

\plotone{fig1.eps}

\clearpage

\plotone{fig2.eps}

\end{document}